\documentstyle[11pt]{article}
\font\cero=cmbx10 scaled 1728
\font\uno=cmcsc10 scaled 1200
\font\dos=cmti10 scaled 1200

\title{{\cero Isometries and embedding of the thermodynamic phase 
space of an ideal gas}} 
\author{{\uno M. Montesinos-Vel\'{a}squez}\thanks{Electronic address : 
merced@fis.cinvestav.mx} \\ 
{\dos Departamento de F\'{\i}sica} \\
{\dos Centro de Investigaci\'{o}n y de Estudios Avanzados del I.P.N.} \\ 
{\dos Apartado Postal 14-740, 07000, M\'{e}xico, D.F., MEXICO.}}
\date{ }
\begin{document}
\maketitle  
\section*{ }
{\uno Abstract}. In this work we find the Killing vector fields of the 
riemannian submanifolds of the thermodynamic phase space of an ideal gas and show that the isometry group corresponding to them is homomorphic to the euclidean group $E(2)$. We also give the embedding of these submanifolds in the euclidean space $(R^{3},\
delta)$.\\[1ex] 
{\uno Resumen}. En este trabajo hallamos los campos vectoriales de 
Killing de las subvariedades riemannianas del espacio fase termodin\'amico de 
un gas ideal y mostramos que el grupo de isometrias correspondiente a cada una de las subvariedades es homeom\'orfico al grupo euclideano $E(2)$. Tambi\'en damos el encaje de estas subvariedades en el espacio euclideano $(R^{3},\delta)$.\\[2ex] 
PACS: 02.40.Ky; 05.70.-a

\section*{\uno 1. Introduction}
Since Weinhold [1--3] found that it is possible to define an intrinsic metric 
structure on a certain vector space associated with each equilibrium state of 
a thermodynamic system [4--7], many efforts have been made to 
understand the implications of the geometrization of thermodynamics 
[3--5,8--10,11]. It has been shown [9] that a riemannian 
metric can be defined on certain submanifolds of the thermodynamic phase space 
whose dimensionality is given by the number of independent intensive variables 
(see also Ref.\ [7]) and that Weinhold's abstract space is isomorphic to 
the tangent space to one of these submanifolds at some equilibrium state. In this paper we adopt the formulation of the equilibrium thermodynamics owing to Weinhold and concretely the version of the Ref.[9]. Contrary to the statement by various authors [6
,12,13,14], this formulation is perfectly valid and it have not inherent contradictions with the classical thermodynamics [15,16]. Moreover, this metric also allows one to define the concept of length for 
fluctuations about equilibrium states [5,10], the physical 
meaning of the metric have been established [4,9] and different 
thermodynamic relations has been derived in Refs.\ [3,5,8,9] employing 
geometrical methods. 

Nevertheless, to establish a correspondence between geometric and 
thermodynamic concepts is an open problem. This work is a contribution in this 
direction. We are interesting in the symmetries and the embedding of the submanifolds of the thermodynamic phase space because it is possible that these symmetries provide us new information about the underlying geometric structure of the thermodynamic ph
ase space. In fact, if we know the Killing vector fields of above submanifolds we know also the integral curves corresponding to them . In this way, it is possible to obtain of a thermodynamic system which evolve throught these integral curves the maximum
 amount of useful work [4,9]. On the other hand, the embedding of the submanifolds of the thermodynamic phase space in other space of bigger dimension permit to know properties of these submanifolds which are not evident with the mathematical tools of the
 intrinsic geometry (characterized by their corresponding metrics). This paper is organizated as follows: In Sect.\ 2 we give the Killing vector fields corresponding to the submanifolds of the thermodynamic phase space of an ideal gas while in Sect.\ 3 we
 
give the embedding of these submanifolds in the euclidean space 
$({R}^3,\delta)$. 

\section*{\uno 2. Killing vector fields} 
The Killing vector fields represent the symmetries of a manifold, consequently 
it is important to find the Killing vector fields of the submanifolds of the thermodynamic phase space of an ideal gas. 

Let $\{x^{\mu}\}$ be local coordinates on $(M,g)$, with $M$ being a riemannian 
manifold with metric tensor $g = g_{\mu\nu}\,dx^{\mu}\otimes dx^{\nu}$ and 
$X=X^{\nu}\,\left(\frac{\partial}{\partial x^{\nu}}\right)$ be a Killing 
vector field on $(M,g)$, then $g_{ \mu\nu}$ and $X^{\nu}$ satisfy the Killing 
equation 
\begin{eqnarray} 
X^{\lambda}{\partial}_{\lambda} g_{\mu\nu} + 
g_{\lambda\nu} {\partial}_{\mu} X^{\lambda} + g_{\mu\lambda} {\partial}_{\nu} 
X^{\lambda} = 0.  
\end{eqnarray}

The submanifolds $(M_{1}, g_{1})$ and $(M_{2},g_{2})$ of the thermodynamic 
phase space of an ideal gas defined by mole number $N = {\rm const.}$ and 
volume $V = {\rm const.}$ have the metrics [7,9] 
\begin{eqnarray} 
g_{1} = \frac{3NR}{2T} \,dT^2 + \frac{NRT}{V^2}\, dV^2, \quad\quad 
(N=\mbox{const.})\\ 
g_{2} = \frac{3NR}{2T} \,dT^2 + \frac{RT}{N}\, 
dN^2, \quad\quad (V=\mbox{const.}). 
\end{eqnarray}

where $T$ is the absolute temperatue and $R$ is the gases universal 
constant. Since the metrics (2)--(3) are flat [7,9], the corresponding 
submanifolds $M_1$ and $M_2$ are maximally symmetric spaces [17], {\em i.e.}, 
they have $2(2+1)/2 = 3$ Killing vector fields. 

By considering the metric (2) and expressing the Killing vector field 
$X$ as $X= X^{T}\left(\frac{\partial}{\partial 
T}\right)+X^{V}\left(\frac{\partial}{\partial V}\right)$, the Killing equation 
reads 
\begin{eqnarray} 
2\left(\frac{3NR}{2T}\right)\, \partial_{T} 
X^{T}- \left(\frac{3NR}{2T^2}\right)\, X^T \!\! & = & \!\! 0, \\ 
2 \left(\frac{NRT}{V^2}\right)\, \partial_{V} X^{V}+ \left(\frac{NR}{V^2}
\right)\, X^T - \left(\frac{2NRT}{V^3}\right)\,X^{V}\!\! & = & \!\! 
0,
\\ \left(\frac{NRT}{V^2}\right)\, \partial_{T} X^{V}+ 
\left(\frac{3NR}{2T}\right)\,\partial_{V} X^T\!\! & = & \!\! 0. 
\end{eqnarray}

A set of independent Killing vector fields that satisfy 
(4)--(6) is given by 
\begin{eqnarray} 
X_{1} & = & 3\sqrt{\frac23} V 
\left(\frac{\partial}{\partial V}\right),\\ 
X_{2} & = & \sqrt{\frac{2T}{3}}\sin{\left(\frac13\sqrt{\frac32}\ln{V}\right)} 
\left(\frac{\partial}{\partial T}\right)\nonumber\\ &&\mbox{}+ 
\frac{V}{\sqrt{T}} \cos{\left(\frac13\sqrt{\frac32}\ln{V}\right)} 
\left(\frac{\partial}{\partial V}\right),\\ 
X_{3} & = & - \sqrt{\frac{2T}{3}}\cos{\left(\frac13\sqrt{\frac32}\ln{V}\right)} 
\left(\frac{\partial}{\partial T}\right)\nonumber\\ &&\mbox{}+ 
\frac{V}{\sqrt{T}} \sin{\left(\frac13\sqrt{\frac32}\ln{V}\right)} 
\left(\frac{\partial}{\partial V}\right), 
\end{eqnarray} 
and it is easy to show that they satisfy the following commutation relations 
\begin{eqnarray} 
[X_1,X_2]=-X_3,\quad\quad [X_1,X_3]=X_2,\quad\quad 
[X_2,X_3]=0.  
\end{eqnarray}

In the case of the metric (3), if the Killing vector 
field $X$ is written as $X= X^{T} \left( \frac{\partial}{\partial T} \right) + 
X^{N} \left( \frac{\partial}{\partial N} \right)$, the Killing equation takes 
the following form 
\begin{eqnarray} 
2 \left( \frac{3NR}{2T} \right)\, \partial_{T} X^{T}- \left( \frac{3NR}{2T^2} 
\right)\, X^T + \left( \frac{3R}{2T} \right) \, X^{N}\!\! & = & \!\! 0, \\ 
2 \left( \frac{RT}{N} \right) \, \partial_{N} X^{N} + \left( \frac{R}{N} 
\right) \, X^T - \left( \frac{RT}{N^2} \right) \, X^{N}\!\! & = & \!\! 0, \\ 
\left( \frac{RT}{N} \right) \, \partial_{T} X^{N} + \left( \frac{3NR}{2T} 
\right) \, \partial_{N} X^T \!\! & = & \!\! 0. 
\end{eqnarray}

In this case, a set of independent Killing vector fields that satisfy 
(11)--(13) are given by 
\begin{eqnarray} 
X_{1} & = & \left( \frac65 \right) \sqrt{\frac23} \left[ T \left( 
\frac{\partial}{\partial T} \right) - N \left( \frac{\partial}{\partial N}
\right) \right],\\ 
X_{2} & = &  \sqrt{\frac{2T}{3N}} \sin{ \left( - \frac12 \sqrt{\frac32} \ln {T} 
+ \frac13 \sqrt{\frac32}\ln{N} \right)} \, \left( \frac{\partial}{\partial T} 
\right) \nonumber\\ 
&&\mbox{} + \sqrt{\frac{N}{T}} \cos{ \left(- \frac12 \sqrt{\frac32}\ln {T} +
\frac13 \sqrt{\frac32}\ln{N} \right)} \, 
\left( \frac{\partial}{\partial N} \right),\\ 
X_{3} & = & \sqrt{\frac{2T}{3N}} \cos{ \left( - \frac12 \sqrt{\frac32} \ln {T} 
+ \frac13 \sqrt{\frac32}\ln{N} \right)} \, \left( \frac{\partial}{\partial T}
\right) \nonumber\\ 
&& \mbox{}- \sqrt{\frac{N}{T}} \sin{ \left(- \frac12 \sqrt{\frac32} \ln {T} + 
\frac13 \sqrt{\frac32} \ln{N} \right)}\, \left( \frac{\partial}{\partial N} 
\right), 
\end{eqnarray} 
which obey the relations 
\begin{eqnarray} 
[X_1,X_2]=-X_3,\quad\quad [X_1,X_3]=X_2,\quad\quad [X_2,X_3]=0. 
\end{eqnarray}

That is to say, (10) and (17) show that the 
riemannian submanifolds of the thermodynamic phase space of ideal gas 
($(M_1,g_1)$ and $(M_2,g_2)$; respectively) have homomorphic isometry groups. In fact, the commutation relations (10) and (17) are identical to those for the two-dimensional euclidean group $E(2)$, consisting of rotations around the origin generated by $X
_1$ and translations generated by $X_2$ and $X_3$ in a two-dimensional euclidean space.

\section*{\uno 3. Embedding in $({R}^3,\delta)$} 
The smooth map $f:M 
\longrightarrow N$ ($M$ and $N$ are smooth manifolds) is called an embedding 
if $f$ is an injection (one-to-one) and an immersion, {\em i.e.}, $f_{\ast}:T_p 
M\longrightarrow T_{f(p)} N$ is an injection. In this section we give the 
embedding of the submanifolds $(M_1,g_1)$ and $(M_2,g_2)$ in 
$({R}^3,\delta)$; with $\delta$ being the euclidean metric on ${R}^3$. 

If we consider the map
\begin{eqnarray}
f & : & M_1 \longrightarrow {R}^3\nonumber\\
  &   & (T,V) \longmapsto (f^1(T,V),f^2(T,V),f^3(T,V)), 
\end{eqnarray}
where
\begin{eqnarray}
f^1(T,V) & = & 
\sqrt{3NRT}\,\cos{\left(\frac13\sqrt{\frac32}\ln{V}\right)},\\ 
f^2(T,V) & = & \sqrt{3NRT}\,\cos{\left(\frac13\sqrt{\frac32}\ln{V}\right)},\\ 
f^3(T,V) & = & \sqrt{6NRT}\,\sin{\left(\frac13\sqrt{\frac32}\ln{V}\right)},
\end{eqnarray} 
then it is clear that (18)--(21) is an 
embedding of $(M_1,g_1)$ in $({R}^3,\delta)$ provided that 
$\frac13\sqrt{\frac32}\ln{V}$ be restricted to a period of the sine and cosine 
functions, for example $[0,2\pi]$ in order to make the map $f$ one-to-one.

On the other hand, the map
\begin{eqnarray}
h & : & M_2 \longrightarrow {R}^3\nonumber\\
  &   & (T,N) \longmapsto (h^1(T,N),h^2(T,N),h^3(T,N)), 
\end{eqnarray}
where
\begin{eqnarray}
h^1(T,N) & = & \sqrt{\frac{6RTN}{5}}\,\cos{\left(-
\frac12\sqrt{\frac32}\ln{T}+\frac13\sqrt{\frac32}\ln{N}\right)},\\ 
h^2(T,N) & = & \sqrt{\frac{6RTN}{5}}\,\cos{\left(-
\frac12\sqrt{\frac32}\ln{T}+\frac13\sqrt{\frac32}\ln{N}\right)},\\ 
h^3(T,N) & = & \sqrt{\frac{12RTN}{5}}\,\sin{\left(-
\frac12\sqrt{\frac32}\ln{T}+\frac13\sqrt{\frac32}\ln{N}\right)}, 
\end{eqnarray} 
is an embedding of $(M_2,g_2)$ in $({R}^3,\delta)$ provided that $-
\frac12\sqrt{\frac32}\ln{T}+\frac13\sqrt{\frac32}\ln{N}$ be restricted to 
a period of the sine and cosines functions, for example $[0,2\pi]$ in 
order to make the map $h$ one-to-one. 

It is obvious that the induced metrics on $M_1$ and $M_2$ are $g_1=f^{\ast} 
\delta = (df^1)^2+(df^2)^2+(df^3)^2$ and $g_2=h^{\ast} \delta=(dh^1)^2+(dh^2)^2+(dh^3)^2$; respectively (compare Refs. [13] and [14]). 

\section*{\uno 4. Concluding remarks} 
We have found the Killing vector fields 
corresponding to the submanifolds $(M_1,g_1)$ and $(M_2,g_2)$ of the 
thermodynamic phase space of an ideal gas and we have showed that the 
isometry groups for both $(M_1,g_1)$ and $(M_2,g_2)$ are homomorphic 
to the two-dimensional euclidean group $E(2)$. On the other hand, the physical meaning of the embedding of $(M_1,g_1)$ and $(M_2,g_2)$ in $({R}^3,\delta)$ it is not clear. Nevertheless, it is possible that the study of systems with nonvanishing curvature 
(perhaps, the van der Waals gas [9] or another system) allows one to understand better this situation. I remark that the isometry group (if it exist) for van der Waals gas and another thermodynamic systems has been not established.
\section*{\uno Acknowledgements}
The author is grateful to Dr.\ G.F. Torres del Castillo for useful comments and 
to CONACyT for financial support. 

\section*{\uno References}
\newcounter{ref} \begin{list}{\hspace{1.3ex}\arabic{ref}.\hfill}
{\usecounter{ref} \setlength{\leftmargin}{2em} \setlength{\itemsep}{-.98ex}}
\item F. Weinhold, {\it J. Chem.\ Phys.}\ {\bf 63} (1975) 2479.
\item F. Weinhold, {\it J. Chem.\ Phys.}\ {\bf 63} (1975) 2484.
\item F. Weinhold, {\it J. Chem.\ Phys.}\ {\bf 63} (1975) 2488.
\item P. Salamon, B. Andresen, P.D. Gait and R.S. Berry, {\it J. Chem.\ 
Phys.}\ {\bf 73} (1980) 1001. 
\item R. Gilmore, {\it Catastrophe theory for scientists and engineers}, Wiley, 
New York (1981); Sect.\ 10.12. 
\item R. Gilmore, {\it Phys.\ Rev.}\ {\bf A30} (1984) 1994. 
\item J.D. Nulton and P. Salamon, {\it Phys.\ Rev.}\ {\bf A31} (1985) 2520. 
\item M. Montesinos-Vel\'asquez, Tesis de Licenciatura, UAP, (1992). 
\item G.F. Torres del Castillo and M. Montesinos-Vel\'asquez, {\it Rev.\ Mex.\ 
F\'{\i}s.}\ {\bf 39} (1993) 194. 
\item G. Ruppeiner, {\it Phys.\ Rev.}\ {\bf A20} (1979) 1608; {\bf 24} (1981) 
488; {\bf 27} (1983) 1116; {\it Phys.\ Rev.\ Lett.}\ {\bf 50} (1983) 287.
\item see G. Ruppeiner, {\it Rev. Mod. Phys.} {\bf 67} (1995) 605 and references therein.
\item R. Gilmore, {\it Phys. Rev.} {\bf A32} (1985) 3144.
\item B. Andresen et al, {\it Phys. Rev.} {\bf A37} (1988) 845.
\item B. Andresen et al, {\it Phys. Rev.} {\bf A37} (1988) 849.
\item G. Ruppeiner, {\it Phys. Rev} {\bf A32} (1985) 3141.
\item K. Horn, {\it Phys. Rev. } {\bf A32} (1985) 3142.
\item D. Kramer, H. Stephani, E. Herlt and M. MacCallum, {\it Exact solutions 
of Einstein's field equations}, Cambridge University Press, Cambridge (1980); 
Sect.\ 8.5. 
\end{list}
\end{document}